%% file: long_paper.tex
\documentclass{article}
\usepackage{kr}

\usepackage{times}
\usepackage{soul}
\usepackage{url}
\usepackage[hidelinks]{hyperref}
\usepackage[utf8]{inputenc}
\usepackage[small]{caption}
\usepackage{graphicx}
\usepackage{amsmath}
\usepackage{amsthm}
\usepackage{amsfonts} 
\usepackage{amssymb}
\usepackage{makecell}
\usepackage{multirow}
\usepackage{tikz}
\usetikzlibrary{trees,chains,scopes,arrows,shapes}
\usepackage{booktabs}
\usepackage{algorithm}
\usepackage{algorithmic}
\usepackage{subcaption}
\usepackage{listings}
\usepackage[disable]{todonotes} 
\makeatletter
\if@todonotes@disabled
  \newcommand{\rnewc}[2][]{}
  \newcommand{\roldc}[2][]{}
  \newcommand{\rnew}[1]{#1}
  \newcommand{\rold}[1]{}
\else
  \newcommand{\rnewc}[2][]{\todo[author=Comment,caption={}, #1]{#2}}
  \newcommand{\roldc}[2][]{\todo[author=Comment,caption={},color=green!40, #1]{#2}}
  \newcommand{\rnew}[1]{\textcolor{blue}{#1}}
  \newcommand{\rold}[1]{\textcolor{red}{\st{#1}}}
\fi

\newcommand{\key}[1]{\textbf{#1}}
\newcommand{\ikey}[1]{\emph{#1}}
\newcommand{\q}[1]{`{#1}\mathchar`'}

\title{Argumentative Reasoning with Language Models on Non-factorized Case Bases}

\author{%
Wachara Fungwacharakorn$^1$\and
May Myo Zin$^1$ \and
Ha Thanh Nguyen$^{1,2}$ \and
Yuntao Kong$^1$ \and
Ken Satoh$^1$\\
\affiliations
$^1$ Center of Juris-Informatics, Joint Support-Center for Data Science Research, ROIS, Tokyo, Japan\\
$^2$ Research and Development Center for Large Language Models, NII, ROIS, Tokyo, Japan
\emails
\{wacharaf, maymyozin, nguyenhathanh, kongyt, ksatoh\}@nii.ac.jp
}

\begin{document}

\maketitle

\begin{abstract}



In this paper, we investigate how language models can perform case-based reasoning (CBR) on non-factorized case bases. We introduce a novel framework, argumentative agentic models for case-based reasoning (AAM-CBR), which extends abstract argumentation for case-based reasoning (AA-CBR). Unlike traditional approaches that require\rold{ manual} factorization of previous cases, AAM-CBR leverages language models to determine case coverage and extract factors based on new cases. This enables factor-based reasoning without exposing or preprocessing previous cases, thus improving both flexibility and privacy. We also present initial experiments to assess AAM-CBR performance by comparing the proposed framework with a baseline that uses a single-prompt approach to incorporate both new and previous cases. The experiments are conducted based on a synthetic credit card application dataset. The result shows that AAM-CBR surpasses the baseline \rnew{only} when the new case contains a richer set of factors. The finding indicates that language models\rold{, on their own} \rnew{can handle case-based reasoning with a limited number of factors, but face challenges as the number of factors increase.}\rold{still find case-based reasoning challenging, while} \rnew{Consequently, integrating}\rold{cooperating} symbolic reasoning with language models, as implemented in AAM-CBR, \rold{becomes crucial as the number of factors increases}\rnew{is crucial for effectively handling cases involving many factors}. 


\end{abstract}

\section{Introduction}

\key{Case-based reasoning} (CBR) is a classic reasoning task in artificial intelligence (AI), widely applied in domains such as law, finance, and healthcare. In CBR, decisions for new cases are derived by drawing analogies to previous cases with known outcomes. One major representation of cases in CBR is a set of \ikey{factors}\rold{, pioneered by early legal CBR systems} \cite{aleven1995doing}. Each factor captures a significant dimension that influences the outcome. These factors are \rold{manually} annotated and serve as abstracted features that allow analogical comparisons between cases. Reasoning proceeds by identifying similarities and differences in the factor sets of the current and previous cases, and applying heuristics or argumentation approaches to reach the outcome for a new case. 

Although factor-based representations support structured and interpretable reasoning, several limitations remain. One limitation is the impossibility of qualifying all factors in the first place, as new cases can lead to new factors. Another limitation is the significant effort to identify factors in previous cases, as the reasoning usually considers a large number of previous cases, and each previous case is described in a very long text. In addition to that, sometimes it is necessary to identify the magnitudes or dimensions of factors, as they can affect the reasoning.


This paper raises an alternative question: can we use language models for case-based reasoning without initially factorizing previous cases, but the reasons can still be interpretable as factor-based reasoning. To reduce the scope of the question, this paper focuses particularly on \key{abstract argumentation for case-based reasoning} (AA-CBR) \cite{cyras2016abstract}. 
Originally, AA-CBR requires \rold{the consideration of factors in previous cases}\rnew{factorizing previous cases} to determine \rold{the }relevance \rold{of the case}, especially against new cases. However, in this paper, we extend AA-CBR into a novel framework, \key{argumentative agentic models for case-based reasoning} (AAM-CBR), which instead takes advantage of language models to determine \rold{the }relevance \rold{of the case}. Therefore, the proposed framework does not require factorizing previous cases. Only new cases are factorized and used to determine \rold{the }relevance \rold{of the case based on the coverage of the factors. The determination is done} by \rold{an individual} language model agent\rnew{s} attached to \rold{each} previous case\rnew{s}. This benefits from propagating new factors from new cases, as well as keeping the information in the previous case private. In addition, this paper presents initial experiments on whether AAM-CBR, which is partially black-boxed in case coverage determination and case factor extraction, can perform better than putting previous cases and new cases in a single prompt, which is solely black-boxed, in predicting the AA-CBR outcome. The experiments are grounded in the domain of credit card application decisions. The result shows that AAM-CBR performs better than the single-prompt approach only when new cases cover more factors and struggles when new cases cover fewer factors. This highlights the importance of integrating symbolic reasoning with language models, as implemented in AAM-CBR, especially when the number of factors increases.

The paper is structured as follows. Section \ref{sec:related-work} provides related work on the background of case-based reasoning, abstract argumentation, and largelanguage models. Section \ref{sec:aacbr} provides the background of abstract argumentation for case-based reasoning (AA-CBR). Section \ref{sec:framework} presents the proposed framework, argumentative agentic models for case-based reasoning (AAM-CBR). Section \ref{sec:experiment} describes the experiments conducted for this paper. Section \ref{sec:result} presents the results of the experiments. Section \ref{sec:discussion} discusses the results and suggests future work. Finally, Section \ref{sec:conclusion} concludes this paper.

\section{Related Work}
\label{sec:related-work}

\rnew{This section provides the backgrounds of} \rold{The intersection of} argumentative reasoning, case-based reasoning, and large language models\rold{ represents a convergence of several distinct research traditions. This section provides a comprehensive survey of relevant work across these domains}, positioning our AAM-CBR framework within\rold{  the broader landscape of AI reasoning systems} \rnew{those backgrounds}.

\subsection{Argumentation and Case-based Reasoning}
Since the foundation of abstract argumentation framework \cite{dung1995acceptability}, there have been numerous computational argumentation extensions that address practical reasoning scenarios. Dung, Mancarella, and Toni \shortcite{dung2002argumentation} provided proof procedures for credulous and sceptical nonmonotonic reasoning, establishing computational foundations that enable practical reasoning implementations. Bench-Capon et al. \shortcite{bench2003computational} explored the integration of computational argumentation with legal practice, providing insights into how formal argumentation can support real-world legal reasoning. \rold{This work established important connections between theoretical argumentation and practical legal applications.}

 \rnewc[inline]{
Remove Atkinson's works as reviewer found they are not relevant

Atkinson et al. \shortcite{atkinson2006computational} proposed influential models to computationally represent practical arguments, extending to action-based systems for presumptive reasoning \cite{atkinson2007practical}.
 Modgil and Caminada \shortcite{modgil2009proof} developed comprehensive proof theories and algorithms for abstract argumentation frameworks, establishing computational foundations that enable practical implementations. 
}

In particular, several researchers have focused on integrating argumentative reasoning with case-based reasoning (CBR). Prakken et al. \shortcite{prakken2015formalization} formalized argumentation schemes for legal case-based reasoning within the ASPIC+ framework, demonstrating how structured legal reasoning can be systematically captured through argument-based models.
\v{C}yras et al. \shortcite{cyras2016abstract} introduced a formal framework bridging these two paradigms, allowing CBR to leverage the inferential structure and evaluation semantics of abstract argumentation. In addition to this, Al Abdulkarim \shortcite{al2017representation} proposed methods to represent case law in a form suitable for argumentative reasoning, with a focus on legal applications. These contributions collectively underscore the potential of combining computational argumentation and CBR to enhance the explainability, structure, and normative grounding of AI legal reasoning systems  \cite{atkinson2020explanation}. 

The relevance of computational argumentation has been further explored in the context of AI systems that require explanability and contestability. Rotolo and Sartor \shortcite{rotolo2023argumentation} investigated computational argumentation and explanation in law, connecting between formal justification and explanation. Leofante et al. \shortcite{leofante2024contestable} proposed the necessity of computational argumentation for building contestable AI systems, advocating for structured reasoning mechanisms that can justify and challenge AI decisions in complex environments.

\subsection{Neural and Hybrid Case-Based Reasoning}

The integration of neural networks with traditional CBR has been long investigated. Early work by Arditi and Tokdemir \shortcite{arditi1999comparison} conducted empirical comparisons in engineering domains, demonstrating that neural networks excel at pattern recognition while CBR provides better interpretability. Chen and Burrell \shortcite{chen2001case} provided a systematic comparison between case-based reasoning systems and artificial neural networks, identifying complementary strengths that could be leveraged through hybrid approaches. 

Recent advances have focused on deep learning integration. Li et al. \shortcite{li2018deep} introduced prototype-based neural networks that perform case-based reasoning through learned prototypes, achieving both accuracy and interpretability. The work demonstrates how neural networks can learn meaningful case representations while maintaining the explanatory power of traditional CBR. Amin et al. \shortcite{amin2018answering} proposed a hybrid system combining deep neural networks with CBR for complex reasoning tasks, showing how deep learning can enhance case retrieval and adaptation processes. More recently, Gould and Toni \shortcite{gould2025neuro} introduced Gradual Abstract Argumentation for Case-Based Reasoning (Gradual
AA-CBR), a neuro-symbolic model utilizing \rold{ an argumentation structure with neural extracted features}\rnew{case-based reasoning as an end-to-end method}, further advancing the goal of interpretable and structured neural case-based reasoning.

Meanwhile, recent research has explored the potential of large language models (LLMs) in addressing limitations of factor-based reasoning in CBR. Gray et al. 
\shortcite{gray2023automatic}
demonstrated the use of LLMs for automatic identification of relevant factors from textual case descriptions. Subsequent work
\cite{gray2024using} investigated how LLMs can be employed to discover novel factors from case corpora, expanding the representational capacity of traditional approaches, and further research by the same group focused on estimating the magnitudes of such factors using LLMs \cite{gray2025generating}. These efforts highlight the growing role of LLMs in augmenting core CBR processes such as case representation and factor extraction, pushing the boundaries of hybrid reasoning systems.

\subsection{Large Language Models for Reasoning}

The application of large language models to reasoning tasks has emerged as a major research direction. Huang and
Chang \shortcite{huang2022towards} provided an early \rold{comprehensive }survey of reasoning capabilities in LLMs, identifying key challenges and opportunities.  Plaat et al. \shortcite{plaat2024reasoning} offered a more recent perspective, examining how reasoning abilities have evolved with larger and more sophisticated models. 

In domain-specific applications, particularly in law, logical reasoning reliability and interpretability are critical, and LLMs still struggle with such reasoning capabilities. Nguyen et al. \shortcite{nguyen2023well} evaluated how well state-of-the-art legal reasoning models support abductive reasoning, a key capability for forming explanatory hypotheses from observations. Despite strong performance in certain legal tasks, current models still struggle with abductive inference. Similarly, Nguyen et al. \shortcite{nguyen2023negation} assessed negation detection in GPT models and found that even the strongest models\rnew{,} like GPT-4 \rnew{at that time,} face notable challenges. These results underscore that key aspects of logical reasoning, such as negation handling and hypothesis generation, remain unresolved\rold{ and are essential for high-stakes domains like law, healthcare, and science}.

To enhance LLM reasoning capabilities, multi-step reasoning has received particular attention. Aksitov et al. \shortcite{aksitov2023rest} developed self-improvement methods for multi-step reasoning in LLM agents, combining reinforcement learning with trajectory optimization. Wang et al. \shortcite{wang2024q} introduced the Q* framework for improving multi-step reasoning through deliberative planning, addressing pathologies in sequential reasoning processes. 

Argumentative reasoning with LLMs has also attracted growing attention. Castagna, Sassoon, and Parsons \shortcite{castagna2024critical} proposed enhancing LLM reasoning by incorporating critical questions inspired by Toulmin’s model of argumentation \cite{toulmin2003uses}, aiming to improve the coherence and structure of model-generated arguments. Sukpanichnant, Rapberger, and Toni \shortcite{sukpanichnant2024peerarg} presented \ikey{PeerArg}, an argumentative reasoning framework using LLMs to support scientific peer review. Liga, Markovich, and Yu \shortcite{liga2025addressing} introduced a hybrid approach that combines abstract argumentation with LLMs through prompt engineering, specifically designed to address legal requirements such as the right to explanation and the right to challenge. Freedman et al. \shortcite{freedman2025argumentative} introduced argumentative LLMs (\ikey{ArgLLMs}), a method to augment LLMs with argumentative reasoning to enhance the explainability of LLMs and allow users to contest LLMs to correct mistakes. Collectively, these approaches illustrate the growing potential of combining LLMs with computational argumentation to enhance explainability and contestability in generated reasoning.

\section{AA-CBR}
\label{sec:aacbr}

{This section provides the background on \key{abstract argumentation for case-based reasoning} (AA-CBR) \cite{cyras2016abstract}. In AA-CBR, a previous case is considered as a pair of a finite set of factors and their decided outcome. Let $\mathbb{F}$ be a set of all possible factors. Each subset of $\mathbb{F}$ is called a \ikey{situation}. AA-CBR assumes a binary distribution of outcomes, which we assume as a set $\mathbb{O} = \{0,1\}$. A \ikey{previous case} is now a pair of $(X,o) \in 2^\mathbb{F} \times \mathbb{O}$. A \ikey{case base} $\Gamma$ is a finite set of outcome-consistent cases (i.e., for $(X,o_x),(Y,o_y ) \in \Gamma$, if $X = Y$, then $o_x = o_y$).

AA-CBR uses an abstract argumentation framework (AA framework) \cite{dung1995acceptability}, which we recap as follows. AA framework is a pair $(\mathcal{A}, \leadsto)$. Each element of $\mathcal{A}$ represents an \ikey{argument} and $\leadsto$ is a binary relation over $\mathcal{A}$ representing \ikey{attacks} between arguments. For $x,y \in \mathcal{A}$, if $x \leadsto y$ then we say $x$ attacks $y$. For a set of arguments $E \subseteq \mathcal{A}$ and an argument $x 
\in \mathcal{A}$, $E$ defends $x$ if, for every $y \in \mathcal{A}$ that attacks $x$, there is an argument $z \in E$ that attacks $y$. Then, the \emph{grounded extension} of $(\mathcal{A}, \leadsto)$ can be constructed inductively as $G = \bigcup_{i \geq 0} G_i$, where $G_0$ is the set of unattacked arguments, and for $i \geq 0$, $G_{i+1}$ is the set of arguments that $G_{i}$ defends. 

AA-CBR needs to assume a \ikey{default} outcome $o_d \in \mathbb{O}$, which is inferred as an outcome for the empty case. $\bar{o}_d$ represents the contrary one (i.e., $\bar{o}_d \in  \mathbb{O}\setminus\{o_d\}$). Given a case base, an \ikey{AA framework} corresponding to $\Gamma$, a default outcome $o_d \in \mathcal{O}$, and a new case $N \subseteq \mathcal{F}$ is $(\mathcal{A}, \leadsto)$ satisfying the following conditions \cite{cyras2016abstract}:
  \begin{enumerate}
      \item (\ikey{arguments}) $\mathcal{A} = \Gamma \cup \{(N,?)\} \cup \{(\emptyset,o_d)\}$;
      \item (\ikey{case attacks}) for $(X,o_x),(Y,o_y) \in \Gamma \cup \{(\emptyset,o_d)\}$, it holds that $(X,o_x) \leadsto (Y,o_y)$ iff
      \begin{itemize}
          \item (different outcomes) $o_x \neq o_y$ , and 
          \item (specificity) $Y \subsetneq X$, and 
          \item (concision) $\not\exists(Z, o_x) \in \Gamma$ with $Y \subsetneq Z \subsetneq X$; 
      \end{itemize}
      \item (\ikey{irrelevant attacks}) for $(Y,o_y) \in \Gamma$, $(N,?) \leadsto (Y,o_y)$ holds iff $Y \not\subseteq N$.
  \end{enumerate}
  The AA-CBR outcome of the new case $N$ is
  \begin{itemize}
      \item the default outcome $o_d$ if $(\emptyset,o_d)$ is in the grounded extension of the corresponding AA framework;
      \item the contrary $\bar{o}_d$, otherwise.
  \end{itemize}

Throughout this paper, we consider the domain of credit card application decisions. The outcome $\q{0}$ represents the rejection of the credit card application and $\q{1}$ represents the approval of the credit card application. We consider the following set $\mathbb{F}_{credit}$ of these factors:
\begin{enumerate}
\item $p_1$: low debt-to-income ratio
\item $p_2$: long and stable employment history
\item $p_3$: consistent payment history on existing loans
\item $p_4$: significant assets declared
\item $p_5$: positive relationship with the bank
\item $n_1$: high number of recent credit inquiries.
\item $n_2$: missed or late payments history
\item $n_3$: insufficient income
\item $n_4$: limited credit history
\item $n_5$: young age
\end{enumerate}

Factor 1-5 (represented as $p_1$-$p_5$)  are \ikey{positive} factors as they favour the approval of the application. Meanwhile, factor 6-10 (represented as $n_1$-$n_5$) are \ikey{negative} factors as they favour toward the rejection of the application. Please note that some CBR models, such as precedential constraint \cite{horty2004result}, consider these favours in their reasoning approach, but not AA-CBR. We just distinguish them here to make cases realistic \rnew{(for detailed comparison between AA-CBR and precedential constraint, see Paulino-Passos and Toni} \shortcite{paulino2021monotonicity}\rnew{ and Fungwacharakorn et al.} \shortcite{fung2025compatibility}\rnew{ )}.

Suppose we have the following previous cases:

\begin{enumerate}
    \item Case 1: the applicant had a limited credit history ($n_4$). This credit card application was rejected.
    \item Case 2: The applicant had an insufficient income ($n_3$) and a limited credit history ($n_4$), but had a long and stable employment history ($p_2$). This credit card application was approved.
\end{enumerate}

We can then represent these previous cases as the following case base:
\begin{align*}
   \Gamma_1 = \{(\{n_4\},0), (\{n_3,n_4,p_2\},1)\}. 
\end{align*}

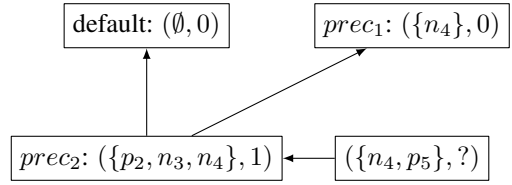
\begin{figure}
    \centering
    \begin{tikzpicture}[>=latex,line join=bevel]
          \node[draw] (n0) at (0bp,100bp) {default: $(\emptyset,0)$};
          \node[draw] (n1) at (100bp,100bp) {$prec_1$: $(\{n_4\},0)$};
          \node[draw] (n2) at (0bp,50bp) {$prec_2$: $(\{p_2,n_3,n_4\},1)$};
          \node[draw] (nq) at (100bp,50bp) {$(\{n_4,p_5\},?)$}; 
          
          \draw [->] (n2)--(n0);
          \draw [->] (n2)--(n1);
          \draw [->] (nq)--(n2);
    \end{tikzpicture}
    \caption{Corresponding AA framework with default outcome $\q{0}$}
    \label{fig:aa0}
\end{figure}

\begin{figure}
    \centering
    \begin{tikzpicture}[>=latex,line join=bevel]
          \node[draw] (n0) at (0bp,100bp) {default: $(\emptyset,1)$};
          \node[draw] (n1) at (0bp,75bp) {$prec_1$: $(\{n_4\},0)$};
          \node[draw] (n2) at (0bp,50bp) {$prec_2$: $(\{p_2,n_3,n_4\},1)$};
          \node[draw] (nq) at (100bp,50bp) {$(\{n_4,p_5\},?)$}; 
          
          \draw [->] (n1)--(n0);
          \draw [->] (n2)--(n1);
          \draw [->] (nq)--(n2);
    \end{tikzpicture}
    \caption{Corresponding AA framework with default outcome $\q{1}$}
    \label{fig:aa1}
\end{figure}
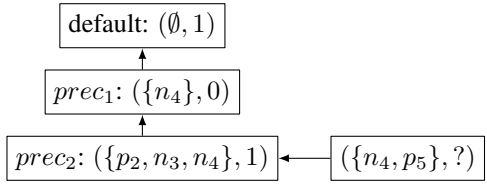

Suppose that we have a new application where the applicant has a limited credit history ($n_4$) but has a positive relationship with the bank ($p_5$). This application can be represented as $N_1 = \{n_4,p_5\}$. If we consider $\q{0}$ as a default outcome, then the AA framework corresponding to $\Gamma_1,\q{0},$ and $N_1$ can be depicted in Figure \ref{fig:aa0}, where $(\emptyset,0)$ is in the grounded extension of the corresponding framework (the grounded extension is $\{(\emptyset,0),(\{n_4\},0), (\{n_4,p_5\},?)\}$). Thus, the AA-CBR outcome of $N_1$ is $\q{0}$. Meanwhile, if we consider $\q{1}$ as a default outcome, then the AA framework corresponding to $\Gamma_1,\q{1},$ and $N_1$ can be depicted in Figure \ref{fig:aa1}, where $(\emptyset,1)$ is not in the grounded extension of the corresponding framework (the grounded extension is $\{(\{n_4\},0), (\{n_4,p_5\},?)\}$). Thus, the AA-CBR outcome of $N_1$ is also $\q{0}$, meaning that the new application should be rejected\rold{, regardless of the default outcome. However, for some case bases and new cases, the AA-CBR outcome can vary depending on the default outcome}.

\section{Proposed Framework}
\label{sec:framework}
This section introduces \key{argumentative agentic models for case-based reasoning} (AAM-CBR), which enables factor-based reasoning on non-factorized case bases. AAM-CBR is designed to operate within the abstract argumentation semantics of AA-CBR while minimizing the need for \rold{manual} preprocessing of previous cases. To achieve this, the framework orchestrates language model agents, each of which is assigned to \rold{each}\rnew{one} previous case. Figure \ref{fig:overview} provides an overview of the AAM-CBR architecture. The reasoning pipeline consists of three primary components: (1) case coverage determination, (2) case factor extraction, and (3) AA-CBR based outcome prediction.
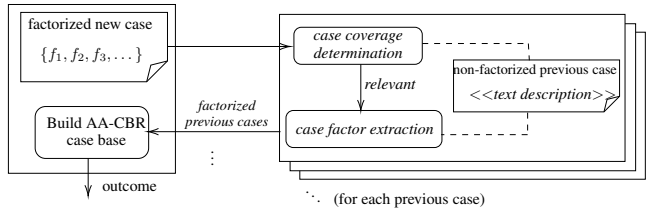
\begin{figure}[ht]
    \centering
    \scalebox{0.5}{
        \input{diagram}
    }
    \caption{Overview of AAM-CBR}
    \label{fig:overview}
\end{figure}

Given a new case represented as a structured set of factors, the AAM-CBR framework determines the appropriate outcome by using a collection of previous cases provided in unstructured natural language form\rold{. These previous cases are stored} without prior factorization\rold{, which preserves privacy and eliminates the need for manual annotation}. 

\rnew{For each agent attached to a previous case, the}\rold{The} process begins with the \ikey{case coverage determination} module, where a language model is used to determine whether the factors of the new case cover the situation described in the textual description\rold{s of previous cases}. If the factors cover the situation, then the previous case is determined as \ikey{relevant} according to AA-CBR. This is similar to \ikey{case coverage} in factor-based reasoning \cite{aleven1995doing,sartor2002teleological}. For the previous case that is deemed ``\textit{relevant}", \rnew{the agent} will proceed to the \ikey{case factor extraction} module. In this module, a language model is used to determine which factors in the new case are implied in the situation of the previous case. The output of the second module is a factorized previous case, composed of factors from the new case implied by the situation in the previous case. \rnew{For the previous case that is not deemed relevant, however, the agent will refuse the response from the new case agent in order to maintain privacy.} This is performed in parallel for every previous case in the case base. The factorized previous cases are then used to build an AA-CBR case base to predict an outcome. Therefore, if the case coverage determination and the case factor extraction work perfectly, the framework will predict the same outcome as AA-CBR.

\section{Experimental Setup}
\label{sec:experiment}

This section describes the experiments used in this paper. The experiments are grounded in the domain of credit card application decisions, based on the same set $\mathbb{F}_{credit}$ of factors shown in Section \ref{sec:aacbr}. The experiments include the steps of generating scenarios, generating test sets, determining case coverage, extracting case factors, and predicting the outcome. The details of each step are as follows. 

\subsection{Generating Scenarios}
\label{subsec:scenario}
The first step is to enumerate all subsets of $\mathbb{F}_{credit}$ and generate scenarios. A scenario is defined as \rold{a textual description that considers only}\rnew{an example textual description of situation that can be represented by} the factors in the subset under consideration and not those outside the subset. To generate a scenario, we use the following prompt template:

\begin{lstlisting}[caption={Generating scenarios}]
TASK: 
Your task is to generate an example of credit card application scenario that covers a specified set of factors and excludes another specified set of factors.
INSTRUCTIONS: 
You will be provided with a specified set of factors that should be covered in the generated scenario and another specified set of factors that should NOT be covered in the generated scenario.
The set of factors that should be covered in the generated scenario:
    {included_factor_list}
The set of factors that should NOT be covered in the generated scenario:
    {excluded_factor_list}
OUTPUT FORMATTING: Generate the scenario in one concise description. Do NOT explicitly use the same words as those in factors. Do NOT include an outcome whether the credit card is accepted or rejected.
\end{lstlisting}

To check whether the scenario considers only the factors in the subset, a language model is then used to extract the factors back from the scenario description, with the following prompt template:

\begin{lstlisting}[caption={Extracting factors}]
TASK: 
Your task is to extract factors from a description of a credit card application scenario.
INSTRUCTIONS: 
You will be provided with a description of a credit card application scenario and a list of all possible factor sentences.
Description: {description}
All possible factors: {all_factor_sentences}
Identify and return ONLY the factor sentences from the provided list that are explicitly present or clearly implied in the description.
OUTPUT FORMATTING: a JSON array of the extracted factor sentences. If no factors are found, return [].
\end{lstlisting}

If the set of extracted factors is the same as the original subset, then we keep that scenario. Otherwise, we tried to generate a new scenario corresponding to that subset upto 10 times. If it still fails after 10 times, then we just skip that subset. Since $\mathbb{F}_{credit}$ has 10 factors, $2^{10} - 1 = 1023$ scenarios can be generated at most from the non-empty subsets of $\mathbb{F}_{credit}$. However, since some subsets are skipped, the number of actual scenarios is slightly lower than 1023. Here is one example of generated scenario, for a subset $\{p_1,n_2\}$:

\begin{quote}
Sarah, a middle-aged individual, applies for a new credit card. Her monthly expenses are manageable compared to her earnings, resulting in a favorable debt-to-earnings ratio. However, her record includes a few instances of overdue bills from years past. While not excessively burdensome, these past indiscretions are recorded on her credit report. She is applying to a new financial institution.
\end{quote}

\subsection{Generating Test Sets}
\label{subsec:testset}
\rnew{Next, we generate test sets based on the generated scenario.} Each test set contains 10 assumed previous cases and 5 new cases. Each previous case is randomly selected from the scenarios and assigned either an outcome $\q{0}$ (the credit card application was rejected) or $\q{1}$ (the credit card application was approved), with the following constraints:

\begin{enumerate}
    \item If the selected scenario is generated from the subset with merely negative factors (resp. positive factors), then the outcome must be $\q{0}$ (resp. $\q{1}$).
    \item To maintain the outcome consistency, if the scenario has already been selected as a previous case in the same test set, then it is assigned with the same outcome.
\end{enumerate}

Meanwhile, a new case is merely a subset of $\mathbb{F}_{credit}$ with no outcome assigned. Each test set contains 5 new cases, each having 6, 7, 8, 9 and 10 factors (the 10-factor new case is then unique and identical to $\mathbb{F}_{credit}$).

\subsection{Determining Case Coverage}

Then, we conducted an experiment on the first module of AAM-CBR: case coverage determination. The experiment considers each pair of previous and new cases (hence, $10 \times 5 = 50$ pairs in each test set). 
Each pair is considered whether or not the factors in the new case cover the situation in previous case, which is counted as \ikey{relevant} in AA-CBR. Unlike the original AA-CBR, we instruct the LLM to determine case coverage by the case description from the previous case and the factors of the new case instead, with the following prompt template:

\begin{lstlisting}[caption={Determing case coverage}]
TASK: 
Your task is to determine whether the factor list covers the case.
INSTRUCTIONS: 
You will be provided with a factor list and a case description.
Here is the factor list: 
    {factor_list}
Here is the case description: 
    {case_description}
Answer this question: does the factor list cover the case?
OUTPUT FORMATTING: 'YES' or 'NO'
\end{lstlisting}

\subsection{Extracting Case Factors}

After that, we conducted an experiment on the second module of AAM-CBR: case factor extraction. The experiment considers each pair of previous and new cases that have been determined as \ikey{relevant} by the first module. \rold{Since we have already conducted a similar task in the first step, we reuse the same prompt template described in Section 5.1.} \rnew{In this experiment, we reuse Prompt 2 to extract factors}. Hence, the experiment becomes a stability evaluation of the case factor extraction prompt. The difference is that \rold{the call in the first step provides}\rnew{the prompt in the scenario generation is fed by} the list of all factors in $\mathbb{F}_{credit}$ while \rold{the call in the second step provides only}\rnew{the prompt in this experiment is fed by} the factors in the new case (which should return similar results if the previous case is actually relevant to the new case i.e., the factors in the new case cover the previous case). 

\subsection{Predicting Outcomes}
After that, we conducted the main experiment, that is, to evaluate the performance of AAM-CBR in predicting the AA-CBR outcome. Since AA-CBR needs to assume a default outcome, we divide the gold standard for the prediction into two modes, one for the default outcome $\q{0}$ and another for the default outcome $\q{1}$. The prediction considers the set of all previous cases in each test set as a case base and considers an individual new case in each test set. We use a baseline method, called a \ikey{SinglePrompt} prediction, which involves feeding all previous cases and new cases into a single prompt to predict an outcome. A \ikey{SinglePrompt} prediction is divided into two options according to it use of argumentative reasoning structure. The first option, referred to as \ikey{non-instructed}, does not provide any argumentative reasoning structure, as the following prompt template:

\begin{lstlisting}[caption={Predicting outcome (non-instructed)}]
TASK: 
You are an expert Case-Based Reasoning (CBR) system. Your task is to predict the outcome for a new case based on given previous cases.
INSTRUCTIONS: 
You will be provided with previous cases, a new case, and a default outcome.
Here are the previous cases you will be working with.
    {previous_case_list}
And this is the new case to analyze:
    {new_case_list}
The default outcome is '{default_outcome}'
Based on the previous cases provided, what is the most likely outcome for this new case?
\end{lstlisting}

The second option, referred to as \ikey{instructed}, provides the argumentative approach based on the dispute trees in AA-CBR. It extends the previous prompt template by replacing the last paragraph in the template with the following instructions:

\begin{lstlisting}[caption={Predicting outcome (instructed-revision)}]
Please do NOT consider the positiveness and negativeness of factors.
You will use a dialectical process between a proponent and an opponent.
Your decision-making process should follow these steps:
1. The proponent starts by asserting a default claim with empty factors and the default outcome '{default_outcome}'
2. The opponent can challenge the proponent's claim if they can identify a previous case that meets the following criteria:
    - with all factors covered by the new case
    - with the outcome '{opponent_outcome}'
    - If the challenging previous case's factors are NOT covered by the proponent's claimed case, then the opponent CANNOT challenge with this previous case (this is a STRICT condition).
3. The proponent can defend against the opponent's rebuttal if they can identify a previous case that meets the following criteria:
  - with all factors covered by the new case
  - with the outcome '{default_outcome}'
  - If the defending previous case's factors are NOT covered by the opponent's claimed case, then the proponent CANNOT defend with this previous case  (this is a STRICT condition).
4. After considering all possible argumentative paths:
  - If the proponent cannot uphold their initial claim through this process (meaning there is no winning path for them), then the predicted outcome for the new case will be '{opponent_outcome}'
  - Otherwise, the predicted outcome for the new case will be  '{default_outcome}'
According to the steps provided, what is the predicted outcome for the new case?
\end{lstlisting}

We found that the prediction works better if we allow the language model to respond with an  explanation (see Section \ref{sec:discussion}). Therefore, we introduce two-step prompting by letting a language model respond with an explanation first, then extracting the prediction from the first response using the following prompt template:

\begin{lstlisting}[caption={Concluding predicted outcome}]
TASK: Your task is to conclude the predicted outcome from the response.
INSTRUCTIONS: Here is the response:
{first_response}
What is the predicted outcome from this response (answer 'mixed' if the predicted outcome cannot be concluded)?
OUTPUT FORMATTING: '{outcome0}' or '{outcome1}' or 'mixed'.
\end{lstlisting}

For AAM-CBR, we use the results from the second module: case factor extraction, to build an AA-CBR case base to predict the AA-CBR outcome. Therefore, we do not need to instruct a language model additionally.

\section{Experimental Result}
\label{sec:result}

This section presents the results from the  initial experiments to evaluate the AAM-CBR framework based on a synthetic dataset that simulates credit card application decisions. The experiments focus on three core components: case coverage determination, case factor extraction, and outcome prediction. To facilitate comparative evaluation, we assess the performance of AAM-CBR against two single-prompt baseline models, which vary in their use of argumentative reasoning structure. Using 50 synthetic test sets, evaluations were conducted on each test set, which contains 10 previous cases and 5 new cases of increasing factor richness ( $n = 6$ to $n = 10$). We used both GEMINI-2.0-FLASH-LITE and GPT-4o as the underlying language models across all tasks, applying identical inputs and prompts to ensure a fair comparison.

\begin{figure*}
    \centering
    \begin{subfigure}{0.48\textwidth}
        \includegraphics[width=\linewidth]{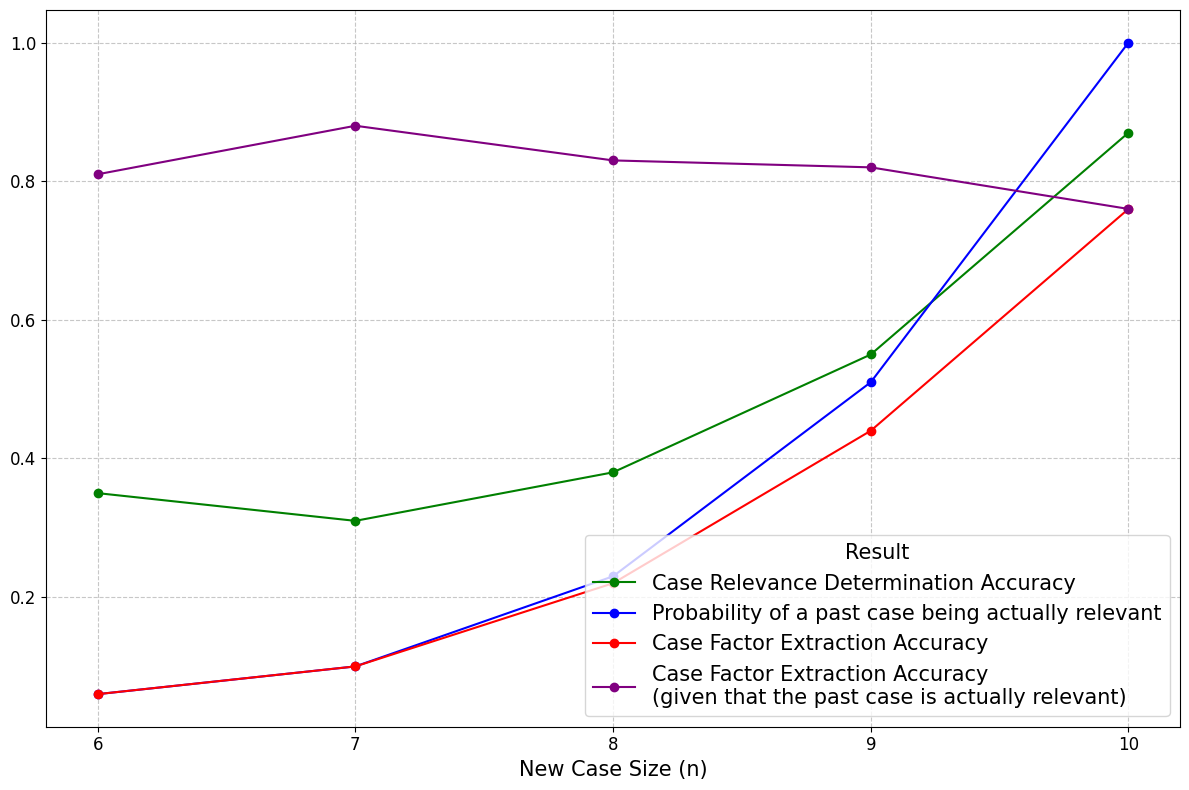}
        \caption{GEMINI-2.0-FLASH-LITE}
        \label{fig:plot}
    \end{subfigure}
    \hfill
    \begin{subfigure}{0.48\textwidth}
        \includegraphics[width=\linewidth]{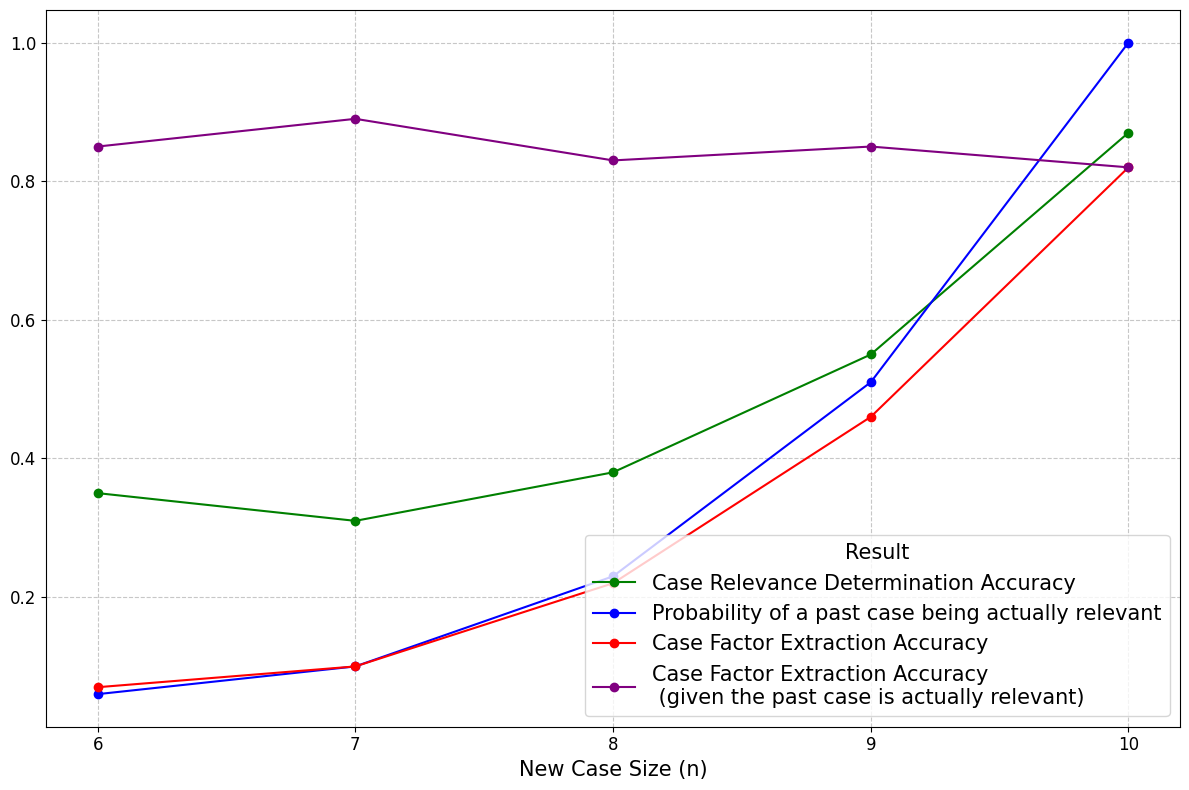}
        \caption{GPT-4o}
        \label{fig:plot_gpt-4o}
    \end{subfigure}
    \caption{Comparative performance on case relevance determination and case factor extraction tasks}
    \label{fig:plot_results}
\end{figure*}

Our first experiment focuses on case relevance determination and factor extraction, varying the number of factors ($n$) in each new case from 6 to 10. As shown in Figure \ref{fig:plot_results}, we report four metrics: (1) \textit{Case Relevance Determination Accuracy}, (2) \textit{Probability that a Retrieved Case is Actually Relevant}, (3) \textit{Case Factor Extraction Accuracy}, and (4) \textit{Factor Extraction Accuracy Given Relevance}. Both models show steady improvements as $n$ increases, with the most significant gains between $n$ = $9$ and $n$ = $10$. While performance trends are closely aligned, GPT-4o consistently achieves slightly higher accuracy in factor extraction, particularly when conditioned on relevance. These results suggest a modest but consistent advantage for GPT-4o in retrieval and structured information extraction tasks.

We then evaluated the final predicted outcomes for each new case using AAM-CBR and two single-prompt baselines. The baselines differ in whether they were explicitly guided to use argumentative reasoning:
\begin{itemize}
\item SinglePrompt-NotInstructed: No guidance on argumentative structure was provided.

\item SinglePrompt-Instructed: Includes explicit prompting for dispute-tree style argumentation.

\item AAM-CBR: Uses structured, dynamically factorized previous cases based on AA-CBR reasoning.
\end{itemize}

Tables \ref{tab:GEMINIresult} and \ref{tab:GPTresult} show the outcome prediction accuracy for GEMINI-2.0-FLASH-LITE and GPT-4o, respectively. Each table shows prediction accuracy split by the new case size ($n$) and the default outcome label (`$0$' or `$1$').

In both models, AAM-CBR underperforms the single-prompt baselines at smaller new cases. For example, in Table \ref{tab:GEMINIresult}, SinglePrompt-NotInstructed achieves an accuracy of 0.76 for outcome `$0$' at $n$ = $6$, while AAM-CBR achieves only 0.28. However, AAM-CBR's performance improves significantly as the number of factors increases. At $n$ = $10$, AAM-CBR achieves 1.00 for outcome `$0$' and 0.96 for outcome `$1$', outperforming both single-prompt baselines by a wide margin.

A similar pattern is observed in Table \ref{tab:GPTresult} for GPT-4o. While the single-prompt baselines initially perform better at $n$ = $6$, AAM-CBR surpasses them from $n$ = $8$ onward. By $n$ = $10$, AAM-CBR achieves 0.98 for outcome `$0$' and 0.96 for outcome `$1$', whereas the best baseline reaches only 0.80 for outcome `$1$'.

These results indicate that structured retrieval and reasoning, as implemented in AAM-CBR, become increasingly important as new cases contain more factors. In contrast, single-prompt methods struggle to maintain performance as the number of factors increases.

\begin{table*}[ht]
    \centering
    \caption{Outcome Prediction Accuracy ( GEMINI-2.0-FLASH-LITE)}
    \label{tab:GEMINIresult}
    \begin{tabular}{|l|c|c|c|c|c|c|c|c|c|c|}
        \hline
        \makecell[c]{\textbf{New case size (n)}} & \multicolumn{2}{c|}{\textbf{n = 6}} & \multicolumn{2}{c|}{\textbf{n = 7}} & \multicolumn{2}{c|}{\textbf{n = 8}}& \multicolumn{2}{c|}{\textbf{n = 9}} & \multicolumn{2}{c|}{\textbf{n = 10}} \\
        \cline{2-11} \makecell[c]{
        $\q{\text{default outcome}}$}
        & \textbf{$\q{0}$} & \textbf{$\q{1}$} & \textbf{$\q{0}$} & \textbf{$\q{1}$} & \textbf{$\q{0}$} & \textbf{$\q{1}$} & \textbf{$\q{0}$} & \textbf{$\q{1}$} & \textbf{$\q{0}$} & \textbf{$\q{1}$} \\
        \hline
        SinglePrompt-NotInstructed & \textbf{0.76} &  0.42 & 0.62 & 0.56 & 0.34 & 0.66 & 0.14 & 0.78 & 0.00 & 0.90 \\
        SinglePrompt-Instructed & 0.56 & \textbf{0.58} & \textbf{0.64} & \textbf{0.68} & 0.44 & 0.54 & 0.40 & 0.36 & 0.40 & 0.48 \\
        AAM-CBR & 0.28 & 0.40 & 0.42 & 0.40 & \textbf{0.70} & \textbf{0.62} & \textbf{0.88} & \textbf{0.82} & \textbf{1.00} & \textbf{0.96} \\
        \hline
    \end{tabular}
\end{table*}

\begin{table*}
    \centering
    \caption{Outcome Prediction Accuracy (GPT-4o)}
    \label{tab:GPTresult}
    \begin{tabular}{|l|c|c|c|c|c|c|c|c|c|c|}
        \hline
        \makecell[c]{\textbf{New case size (n)}} & \multicolumn{2}{c|}{\textbf{n = 6}} & \multicolumn{2}{c|}{\textbf{n = 7}} & \multicolumn{2}{c|}{\textbf{n = 8}}& \multicolumn{2}{c|}{\textbf{n = 9}} & \multicolumn{2}{c|}{\textbf{n = 10}} \\
        \cline{2-11} \makecell[c]{
        $\q{\text{default outcome}}$}
        & \textbf{$\q{0}$} & \textbf{$\q{1}$} & \textbf{$\q{0}$} & \textbf{$\q{1}$} & \textbf{$\q{0}$} & \textbf{$\q{1}$} & \textbf{$\q{0}$} & \textbf{$\q{1}$} & \textbf{$\q{0}$} & \textbf{$\q{1}$} \\
        \hline
        SinglePrompt-NotInstructed & 0.54 &  \textbf{0.60} & 0.56 & \textbf{0.76} & 0.46 & 0.62 & 0.36 & 0.60 & 0.18 & 0.80 \\
        SinglePrompt-Instructed & \textbf{0.64} & 0.54 & \textbf{0.66} & 0.64 & 0.52 & 0.48 & 0.22 & 0.30 & 0.28 & 0.44 \\
        AAM-CBR & 0.24 & 0.46 & 0.42 & 0.48 & \textbf{0.74} & \textbf{0.66} & \textbf{0.84} & \textbf{0.84} & \textbf{0.98} & \textbf{0.96} \\
        \hline
    \end{tabular}
\end{table*}

\section{Discussion}
\label{sec:discussion}

This section discusses several findings of this paper. The first finding is that symbolic reasoning gains importance as the number of factors increases. This aligns with other hybrid case-based reasoning research that demonstrates the necessity of a symbolic framework to improve interpretability, especially when reasoning involves complex interactions among a large number of cases. The second finding is that, although AAM-CBR identifies factors only through new cases, increasing the number of known factors still leads to significant improvements in accuracy. This reflects the capability of LLMs to discover new factors, as the problem of case coverage determination can be reduced to the problem of discovering a new factor (i.e., if we cannot discover a new factor, then the given factors already cover the given case). This finding points to a promising direction for future work: developing methods to incrementally learn new factors through a series of cases. 

On the other hand, the experiments also reveal that, without cooperating with symbolic frameworks, large language models still struggle with case-based reasoning. When conducting the experiment, the baseline \ikey{SinglePrompt} occasionally predicts an outcome using alternative approaches, such as counting positive and negative factors, even with the \ikey{instructed} option, where we explicitly instruct LLMs not to consider the positive and negative nature of factors. Furthermore, the experiments also reveal that the predictions without explicit reasoning are less accurate compared to those with explicit reasoning. The finding is consistent with previous studies on multi-step reasoning and leads us to use two-step prompting to increase the accuracy of the prediction and the stability of the language model response. Specifically, the experiment on the \ikey{case coverage determination} module indicates a bias toward deeming previous cases relevant to new cases, with precision decreasing exponentially as the number of factors in the new case decreases. The decrement follows from the probability that a previous case is actually relevant, which can be calculated as the probability that a subset of $\mathbb{F}$ is also a subset of a new case $N \subseteq \mathbb{F}$ (size $n$), which is equal to $2^{(n-||\mathbb{F}||)}$. This probability decreases exponentially. For example, given $||\mathbb{F}|| = 10$ as in our setting, $n = 8$ gives the probability $2^{(8-10)} = 25\%$ while $n = 6$ gives the probability $2^{(6-10)} = 6.25\%$. Furthermore, the experiment on the \ikey{case extraction} module reveals that large language models still produce unstable responses, resulting in approximately $0.80$ accuracy given that the previous case is actually relevant, even when we reuse the same prompt template for both the $\q{\text{generating scenario}}$ and $\q{\text{extracting case factors}}$ steps. Thus, it still suggests that prompt engineering is required to reduce biases and ensure stability in both modules to realize the full potential of AAM-CBR.

\section{Conclusion}
\label{sec:conclusion}

This paper presents argumentative agentic models for case-based reasoning (AAM-CBR), a novel framework that utilizes language models to perform case-based reasoning without requiring \rold{manual} factorization of previous cases. AAM-CBR leverages language models to dynamically determine case coverage and extract factors directly from new cases, thereby enhancing both flexibility and privacy by not exposing previous cases. Our experiments, conducted in the domain of credit card application decisions, focused on evaluating the core modules of AAM-CBR: \ikey{case coverage determination} and \ikey{case factor extraction}, as well as its overall performance in predicting outcomes. The results demonstrated that AAM-CBR significantly outperforms single-prompt baselines when the new case contains a richer set of factors. The findings underscore the importance of integrating symbolic reasoning with large language models, especially when the number of factors increases. The variation in performance due to the richness of factors in new cases highlights two main challenges for future development. The first challenge is to enhance the language model's capability to accurately identify case coverage and extract relevant factors, particularly when new cases are less comprehensive. The second challenge is to improve the overall prediction by discovering and learning factors from a series of cases. 

\section*{Acknowledgments}
This work was supported by the “R\&D Hub Aimed at Ensuring Transparency and Reliability of Generative AI Models” project of the Ministry of Education, Culture, Sports, Science and Technology, the “Strategic Research Projects” grant
from ROIS (Research Organization of Information and Systems), and JSPS KAKENHI Grant Numbers, JP22H00543.

\bibliographystyle{kr}
\bibliography{kr-sample}

\end{document}

%% file: diagram.tex
\tikzset{every picture/.style={line width=0.75pt}, every node/.style={scale=1.2,yshift=-2}} 

\begin{tikzpicture}[x=0.75pt,y=0.75pt,yscale=-1,xscale=1]

\draw  [fill={rgb, 255:red, 255; green, 255; blue, 255 }  ,fill opacity=1 ] (15,3) -- (183.47,3) -- (183.47,173.4) -- (15,173.4) -- cycle ;
\draw  [fill={rgb, 255:red, 255; green, 255; blue, 255 }  ,fill opacity=1 ] (308.94,30.98) -- (657.47,30.98) -- (657.47,179.4) -- (308.94,179.4) -- cycle ;
\draw  [fill={rgb, 255:red, 255; green, 255; blue, 255 }  ,fill opacity=1 ] (299.15,22.42) -- (647.67,22.42) -- (647.67,170.84) -- (299.15,170.84) -- cycle ;
\draw  [fill={rgb, 255:red, 255; green, 255; blue, 255 }  ,fill opacity=1 ] (288.47,13) -- (636.99,13) -- (636.99,161.42) -- (288.47,161.42) -- cycle ;

\draw   (154.47,81) -- (27.47,81) -- (27.47,11) -- (175.47,11) -- (175.47,60) -- cycle -- (154.47,81) ; \draw   (175.47,60) -- (158.67,64.2) -- (154.47,81) ;

\draw   (615.75,111.4) -- (464,111.4) -- (464,59) -- (631.47,59) -- (631.47,95.68) -- cycle -- (615.75,111.4) ; \draw   (631.47,95.68) -- (618.89,98.82) -- (615.75,111.4) ;

\draw   (303,33.48) .. controls (303,29.35) and (306.35,26) .. (310.48,26) -- (424.99,26) .. controls (429.12,26) and (432.47,29.35) .. (432.47,33.48) -- (432.47,55.92) .. controls (432.47,60.05) and (429.12,63.4) .. (424.99,63.4) -- (310.48,63.4) .. controls (306.35,63.4) and (303,60.05) .. (303,55.92) -- cycle ;

\draw   (295,117.48) .. controls (295,113.35) and (298.35,110) .. (302.48,110) -- (444.84,110) .. controls (448.97,110) and (452.32,113.35) .. (452.32,117.48) -- (452.32,139.92) .. controls (452.32,144.05) and (448.97,147.4) .. (444.84,147.4) -- (302.48,147.4) .. controls (298.35,147.4) and (295,144.05) .. (295,139.92) -- cycle ;

\draw    (175.47,45.4) -- (297.47,45.4) ;
\draw [shift={(299.47,45.4)}, rotate = 180] [color={rgb, 255:red, 0; green, 0; blue, 0 }  ][line width=0.75]    (10.93,-3.29) .. controls (6.95,-1.4) and (3.31,-0.3) .. (0,0) .. controls (3.31,0.3) and (6.95,1.4) .. (10.93,3.29)   ;
\draw  [dash pattern={on 4.5pt off 4.5pt}]  (432.47,42.4) -- (523.47,42.4) -- (539.47,42.4) ;
\draw  [dash pattern={on 4.5pt off 4.5pt}]  (538.73,112.21) -- (538.73,134.4) ;
\draw    (370.47,63.4) -- (370.47,108.4) ;
\draw [shift={(370.47,110.4)}, rotate = 270] [color={rgb, 255:red, 0; green, 0; blue, 0 }  ][line width=0.75]    (10.93,-3.29) .. controls (6.95,-1.4) and (3.31,-0.3) .. (0,0) .. controls (3.31,0.3) and (6.95,1.4) .. (10.93,3.29)   ;
\draw    (289.47,131.4) -- (157.47,131.4) ;
\draw [shift={(155.47,131.4)}, rotate = 360] [color={rgb, 255:red, 0; green, 0; blue, 0 }  ][line width=0.75]    (10.93,-3.29) .. controls (6.95,-1.4) and (3.31,-0.3) .. (0,0) .. controls (3.31,0.3) and (6.95,1.4) .. (10.93,3.29)   ;
\draw   (42,116.48) .. controls (42,110.69) and (46.69,106) .. (52.48,106) -- (145.99,106) .. controls (151.77,106) and (156.47,110.69) .. (156.47,116.48) -- (156.47,147.92) .. controls (156.47,153.71) and (151.77,158.4) .. (145.99,158.4) -- (52.48,158.4) .. controls (46.69,158.4) and (42,153.71) .. (42,147.92) -- cycle ;

\draw    (95.47,158.4) -- (95.47,194.4) ;
\draw [shift={(95.47,196.4)}, rotate = 270] [color={rgb, 255:red, 0; green, 0; blue, 0 }  ][line width=0.75]    (10.93,-3.29) .. controls (6.95,-1.4) and (3.31,-0.3) .. (0,0) .. controls (3.31,0.3) and (6.95,1.4) .. (10.93,3.29)   ;
\draw  [dash pattern={on 4.5pt off 4.5pt}]  (538.73,134.4) -- (451.47,134.4) ;
\draw  [dash pattern={on 4.5pt off 4.5pt}]  (539.47,42.4) -- (540.47,59.4) ;

\draw (312,175) node [anchor=north west][inner sep=0.75pt]   [align=left] {$\displaystyle \ddots~~~$(for each previous case)};
\draw (372.47,71.9) node [anchor=north west][inner sep=0.75pt]  [font=\normalsize] [align=left] {\textit{relevant}};
\draw (176,95) node [anchor=north west][inner sep=0.75pt]  [font=\small] [align=left] {\begin{minipage}[lt]{72.22pt}\setlength\topsep{0pt}
\begin{center}
{\footnotesize \textit{factorized\\ previous cases}}
\end{center}

\end{minipage}};


\draw (217,136.4) node [anchor=north west][inner sep=0.75pt]   [align=left] {$\displaystyle \vdots $};
\draw (108,177) node [anchor=north west][inner sep=0.75pt]   [align=left] {outcome};
\draw (46.75,111.48) node [anchor=north west][inner sep=0.75pt]   [align=left] {\begin{minipage}[lt]{69.04pt}\setlength\topsep{0pt}
\begin{center}
Build AA-CBR \\case base
\end{center}

\end{minipage}};
\draw (293,117.48) node [anchor=north west][inner sep=0.75pt]   [align=left] {\begin{minipage}[lt]{99.65pt}\setlength\topsep{0pt}
\begin{center}
\textit{case factor extraction}
\end{center}

\end{minipage}};
\draw (309,27) node [anchor=north west][inner sep=0.75pt]   [align=left] {\begin{minipage}[lt]{71.32pt}\setlength\topsep{0pt}
\begin{center}
\textit{case coverage}\\\textit{determination}
\end{center}

\end{minipage}};
\draw (478,82) node [anchor=north west][inner sep=0.75pt]   [align=center] {$<<$\textit{text description}$>>$};
\draw (462,58.48) node [anchor=north west][inner sep=0.75pt]   [align=center] {\small{non-factorized previous case}};
\draw (45.28,41) node [anchor=north west][inner sep=0.75pt]   [align=left] {$\displaystyle \{f_{1} ,f_{2} ,f_{3} ,\dotsc \}$};
\draw (33,12) node [anchor=north west][inner sep=0.75pt]   [align=center] {factorized new case};

\end{tikzpicture}